\documentstyle[twocolumn,aps,psfig]{revtex}

\input{psfig}

\newcommand{\bold}[1]{\mbox{\boldmath $#1$}}    
\newcommand{\half}{\mbox{${1\over2}$}}          
\newcommand{\MeV}{{\rm MeV}}                    
\newcommand{\fm}{{\rm fm}}                      
\newcommand{\cappa}{{\mbox{\boldmath{\scriptsize{$\kappa$}}}}}
\newcommand{\Cappa}{{\mbox{\boldmath{{$\kappa$}}}}}
\newcommand{\lambd}{{\mbox{\boldmath{\scriptsize{$\lambda$}}}}}
\newcommand{\Lambd}{{\mbox{\boldmath{{$\lambda$}}}}}
\newcommand{\halpha}{{\hat{\alpha}}}            
\newcommand{\hp}{{\hat{p}}}                     
\newcommand{\hq}{{\hat{q}}}                     
\newcommand{\k}{{\bf k}}                        
\newcommand{\K}{{\bf K}}                        
\newcommand{\bfL}{{\bf L}}                      
\newcommand{\x}{{\bf x}}                        
\newcommand{\ti}{t_{\rm i}}                     
\newcommand{\beq}{\begin{equation}}
\newcommand{\eeq}{\end{equation}}
\newcommand{\beqar}{\begin{eqnarray}}
\newcommand{\eeqar}{\end{eqnarray}}

\begin{document}

\title{Quantum Field Treatment of DCC Dynamics}

\author{J\o rgen Randrup}       

\address{Nuclear Science Division, 
Lawrence Berkeley National Laboratory\\
1 Cyclotron Road, Berkeley, California 94720, U.S.A.}

\date{June 30, 2000}

\maketitle
\begin{abstract}

A practical quantum-field treatment is developed for
systems endowed with an effective mass function
depending on both space and time
and a schematic application illustrates
the quantitative importance of quantum fluctuations
in the dynamics of disoriented chiral condensates.
\end{abstract}


\section{Introduction}
\label{introduction}

The prospect of observing manifestations of chiral symmetry restoration
in high-energy nuclear collisions has stimulated 
significant research activity over the past several years
\cite{DCC}.     
In this endeavor,
the most popular theoretical tool has been the linear $\sigma$ model,
which in its simplest form describes an interacting O(4) field
with isoscalar ($\sigma$) and isovector ($\bold{\pi}$) components.
Within this framework,
a number of instructive dynamical simulations
have been carried out at the level of classical fields
for the types of environment expected in high-energy collisions
\cite{Rajagopal:NPB404,Asakawa:PRL74,class,JR:PRL77,JR:PRC61}.
Furthermore,
quantum-field treatments have been developed and applied,
albeit within the mean-field approximation and for uniform environments
\cite{Boyanovsky,Mottola,JR:HIP9}.

Relative to the classical studies,
these latter treatments take account of the ever present quantum fluctuations
and it appears that these may contribute significantly
to the emerging signals \cite{JR:HIP9}.
Therefore,
it is of interest to further develop the quantal treatment
so that more refined scenarios can be addressed dynamically. 
The present paper describes how the quantum-field treatment
may be extended to non-uniform environments
in a manner that allows numerical simulations 
for the types of scenario relevant to the study of
disoriented chiral condensates.
Moreover,
the important role of quantum fluctuations is illustrated
by means of a concrete application.

The highly excited systems formed in high-energy nuclear collisions
expand very fast
and the environments cool down correspondingly rapidly.
As a consequence,
the effective pion mass decreases quite quickly
 from its initially high thermal value towards its free value.
Furthermore,
the non-equilibrium relaxation dynamics of the chiral order parameter
gives rise to a fairly regular oscillatory modulation of the mass
with a frequency near half the value employed for the $\sigma$ meson
\cite{JR:PRL77}.
Since the expansion is predominantly longitudinal,
it appears \cite{Asakawa:PRL74,JR:PRL77}
that the cooling rate is insufficient to produce a quench
\cite{Rajagopal:NPB404},
in which the square of the pion mass turns negative,
causing the softest modes to become unstable
and thus display an exponential population growth.
While perhaps unfortunate from the physics perspective,
this feature does simplify the physical analysis
and improves the prospect for developing test-particle transport treatments
\cite{JR:PLB435}.
(A concrete impression of the typical time evolution
of the effective pion mass in the bulk of the rapidly expanding system
can be gained from Fig.~\ref{f:mu0}.)

A further complication encountered in nuclear collision experiments
is the non-uniformity of the environment
resulting from the finite extension of the systems formed.
In the mean-field framework,
this causes the mass to be dependent on position as well,
which then couples the elementary modes
and thus complicates the treatment significantly.
(A typical spatial profile of the mass function
expected in nuclear collisions is shown in Fig.~\ref{f:mux}.)

\section{Propagation}
\label{dynamics}

With such environments in mind,
we now turn to the formal developments.
In the mean-field approximation,
the Hamiltonian density operator is given by
\beq
\hat{\cal H}(\x)=
\half[\hp(\x)^2+(\nabla\hq(\x))^2+\mu^2(\x,t)\hq(\x)^2]\ ,
\eeq
where $\hq(\x)$ is the field operator 
and $\hp(\x)$ is its time derivative.
Free pions provide a natural basis
but, in order to bring the problem onto a canonical form,
each pair of traveling waves $|\pm\k\rangle$
is replaced by a corresponding pair of standing waves $|\pm\Cappa\rangle$
\cite{JR:HIP9}.
With an arbitrary sign convention for the quantum numbers $\Cappa=\pm\k$,
the associated canonical eigenfunctions are
\beq\label{xkappa}
\langle\x|\Cappa\rangle\ =\ {1\over\sqrt{\Omega}}\times \left\{
\begin{array}{clll} 
\phantom{-} \sqrt{2}\ \cos(\Cappa\cdot\x) &, & \Cappa> \bold{0} &,\\ 
 1 &, & \Cappa=\bold{0} &,\\
- \sqrt{2}\ \sin(\Cappa\cdot\x) &, & \Cappa<\bold{0} &. \end{array}
\right.
\eeq
The canonical modes are orthonormal and complete,
$\int d\x \langle\Cappa|\x\rangle\langle\x|\Cappa'\rangle
=\delta_{\cappa\cappa'}$,
$\sum_\cappa \langle\x|\Cappa\rangle\langle\Cappa|\x'\rangle=\delta(\x-\x')$.
The canonical annihilation operator is $\halpha^{}_\cappa$,
with $[\halpha^{}_\cappa,\halpha^\dagger_{\cappa'}]=\delta_{\cappa\cappa'}$,
and the dispersion relation is $\omega_\kappa^2=\kappa^2+m^2$.

The field operators can then be expanded as follows,
\beqar\label{phi-psi}
\hq(\x) &=& \sum_\cappa \langle\x|\Cappa\rangle\ \hq_\cappa
= \sum_\lambd \left[ 
F^\lambd(\x) \halpha^{}_\lambd + F^\lambd(\x)^* \halpha^\dagger_\lambd 
\right]\ ,\\
\hp(\x) &=& \sum_\cappa \langle\x|\Cappa\rangle\ \hp_\cappa
=\sum_\lambd \left[ 
G^\lambd(\x) \halpha^{}_\lambd + G^\lambd(\x)^* \halpha^\dagger_\lambd 
\right]\ .
\eeqar
The mode functions are governed by the usual field equation of motion, 
which can readily be solved numerically,
\beq\label{EoM}
\dot{F}^\lambd(\x)\ =\ G^\lambd(\x)\ ,\,\,\
\dot{G}^\lambd(\x)\ =\ [\Delta-\mu^2(\x)]F^\lambd(\x)\ .
\eeq
The appropriate initial conditions are
\beq
F^\lambd(\x,\ti) = 
{1\over\sqrt{2{\omega}_\lambd}} \langle\x|\Lambd\rangle ,\,
G^\lambd(\x,\ti) = 
-i\sqrt{{\omega}_\lambd\over2} \langle\x|\Lambd\rangle .
\eeq
Using the canonial representation of the mode functions,
\beq
F^\lambd(\x)=\sum_\cappa \langle\x|\Cappa\rangle\ F^\lambd_\cappa ,\,\,\,
G^\lambd(\x)=\sum_\cappa \langle\x|\Cappa\rangle\ G^\lambd_\cappa ,
\eeq
we thus obtain the evolution of the canonical operators,
\beqar\label{qt}
\hq_\cappa(t) &=& \sum_\lambd
\left[  F^\lambd_\cappa(t)\ \halpha^{}_\lambd\ 
+\      F^\lambd_\cappa(t)^*\  \halpha^\dagger_\lambd \right]\ ,\\ \label{pt}
\hp_\cappa(t) &=& \sum_\lambd
\left[  G^\lambd_\cappa(t)\ \halpha^{}_\lambd\  
+\      G^\lambd_\cappa(t)^*\  \halpha^\dagger_\lambd \right]\ .
\eeqar

\section{Analysis}
\label{qp}

It is instructive to express the Hamiltonian
in terms of the canonical operators $\hp_\cappa$ and $\hq_\cappa$,
\beq\label{H} 
\hat{H}(t) \equiv \int d\x\ \hat{\cal H}(\x) =
\half\sum_\cappa \hp_\cappa^2
+\half\sum_{\cappa\cappa'}
\hq_\cappa M_{\cappa\cappa'}\hq_{\cappa'}\ .
\eeq
Since the mass matrix 
is real and symmetric,
$M_{\cappa\cappa'}(t)=\langle\Cappa|\mu^2(\x,t)-\Delta|\Cappa'\rangle$,
it can be diagonalized at any given time $t$
by a real, unitary matrix $\bold{S}(t)$, leading to
\beq
\hat{H}(t)\ =\ \half \sum_\K\left[
 \hat{P}_\K^2\ +\ \Omega_\K^2\ \hat{Q}_\K^2 \right]\ .
\eeq
Here the correlated canonical field operators are given by
\beq\label{QP}
\hat{Q}_\K = \sum_{\cappa} \hq_{\cappa} S_{\cappa\K}(t)\ ,\,\,\,\
\hat{P}_\K = \sum_{\cappa} \hp_{\cappa} S_{\cappa\K}(t)\ ,
\eeq
and they satisfy $[\hat{Q}_\K,\hat{P}_{\K'}]=i\delta_{\K\K'}$.
At the given time $t$,
the corresponding correlated eigenfunctions are
\beq
\langle\x|\K\rangle=
\sum_\cappa \langle\x|\Cappa\rangle\ S_{\cappa\K}(t)\ .
\eeq

If the eigenvalues are positive, $\Omega_\K^2>0$,
excitations relative to the adiabatic vacuum state
can be defined by means of correlated quasiparticle annihilation operators,
\beq\label{AK}
\hat{A}^{}_\K = \sqrt{\Omega_\K\over2}\ \hat{Q}_\K
        +{i\over\sqrt{2\Omega_\K}}\ \hat{P}_\K\ ,
\eeq
which satisfy $[\hat{A}^{}_\K,\hat{A}^\dagger_{\K'}]=\delta_{\K\K'}$.
The adiabatic vacuum state, $|\bar{0}\rangle$,
is characterized by the absence of quasiparticles,
$\hat{A}^{}_\K|\bar{0}\rangle=0$,
and is generally a complicated many-body state 
when expressed in terms of the free quanta.

It is important to recognize that (\ref{AK})
refers to the Schr\"odinger representation.
The corresponding Heisenberg operator,
$\hat{\cal A}^{}_\K(t)$,
can be obtained by replacing the Schr\"odinger operators
$\hq_{\cappa}$ and $\hp_{\cappa}$ entering in (\ref{QP})
by the corresponding Heisenberg operators given in (\ref{qt}-\ref{pt}),
\beq
\hat{\cal A}_\K(t)=
\sum_\bfL\left[ {\cal U}_\K^\bfL(t)     \hat{A}_\bfL^{}
+               {\cal V}_\K^\bfL(t)^*   \hat{A}_\bfL^\dagger \right]\ ,
\eeq
where $\hat{A}_\bfL$ is the quasiparticle annihilation operator
associated with the initial form of the mass function $\mu^2(\x,\ti)$.
The Bogoliubov coefficients
${\cal U}_\K^\bfL(t)$ and ${\cal V}_\K^\bfL(t)$
can be obtained from the basic mode coefficients
$F^\lambd_\cappa(t)$ and $G^\lambd_\cappa(t)$
by a sequence of elementary transformations.

Since the operators $\hat{\cal A}_\K(t)$
refer to the evolving eigenrepresentation,
they are less suitable for the analysis
which is better performed in the $\x$ and $\k$ representations,
\beqar
\hat{\cal A}^{}(\x,t) &=&
\sum_\bfL\left[ {\cal U}^\bfL(\x,t)     \hat{A}_\bfL^{}
+               {\cal V}^\bfL(\x,t)^*   \hat{A}_\bfL^\dagger \right]\ ,\\
\hat{\cal A}^{}_\k(t) &=&
\sum_\bfL\left[ {\cal U}_\k^\bfL(\x,t)  \hat{A}_\bfL^{}
+               {\cal V}_\k^\bfL(\x,t)^*\hat{A}_\bfL^\dagger \right]\ .
\eeqar
The corresponding Bogoliubov coefficients
are obtained by the appropriate additional transformations,
\beqar
{\cal U}^\bfL(\x;t) &=&
\sum_{\cappa\K}\langle\x|\Cappa\rangle\ S_{\cappa\K}(t)\ {\cal U}^\bfL_\K(t)\ ,
\\
{\cal U}^\bfL_\k(t) &=&
\sum_{\cappa\K}\langle\k|\Cappa\rangle\ S_{\cappa\K}(t)\ {\cal U}^\bfL_\K(t)\ .
\eeqar

With the above preparations,
we are now able to derive expressions for any quasiparticle observable.
Of particular interest is
the spatial correlation function,
\beqar\nonumber
\bar{\rho}(\x,\x';t) &\equiv&
\langle t|\hat{A}^\dagger(\x) \hat{A}^{}(\x') |t\rangle =
\langle\ti|\hat{\cal A}^\dagger(\x,t) \hat{\cal A}^{}(\x',t) |\ti\rangle\\
&=& 
\bar{\rho}^{\rm qu}(\x,\x';t)+\bar{\rho}^{\rm cl}(\x,\x';t)\ .
\eeqar
Here the first term arises from the quantum fluctuations in the initial state
and is given by
\beq
\bar{\rho}^{\rm qu}(\x,\x';t)=
\sum_\bfL {\cal V}^\bfL(\x,t)\ {\cal V}^\bfL(\x',t)^*\ .
\eeq
The second term is caused by the real agitations in the initial state
(which are often referred to as 
the ``classical'' or ``thermal'' field fluctuations).

Analogously,
we may consider the one-body density matrix in the $\k$ representation,
\beqar\noindent
\bar{\rho}_{\k\k'}(t) &\equiv&
        \langle t|\hat{A}^\dagger_\k \hat{A}^{}_{\k'} |t\rangle
 =      \langle\ti|\hat{\cal A}^\dagger_\k(t)\hat{A}^{}_{\k'}(t)|\ti\rangle\\
&=& \bar{\rho}^{\rm qu}_{\k\k'}(t)+\bar{\rho}^{\rm cl}_{\k\k'}(t)\ ,
\eeqar
where the quantal contribution is
\beq
\bar{\rho}^{\rm qu}_{\k\k'}(t)=
\sum_\bfL {\cal V}^\bfL_\k(t)\ {\cal V}^\bfL_{\k'}(t)^*\ .
\eeq

Expressions for observables reflecting two- or many-body correlations 
can also be derived but they become progressively more complicated.
Moreover, the results of course depend on what is being assumed
regarding the corresponding correlations in the initial state.

\section{Illustrative application}
\label{ill}

\begin{figure}[t]
\centerline{\psfig{figure=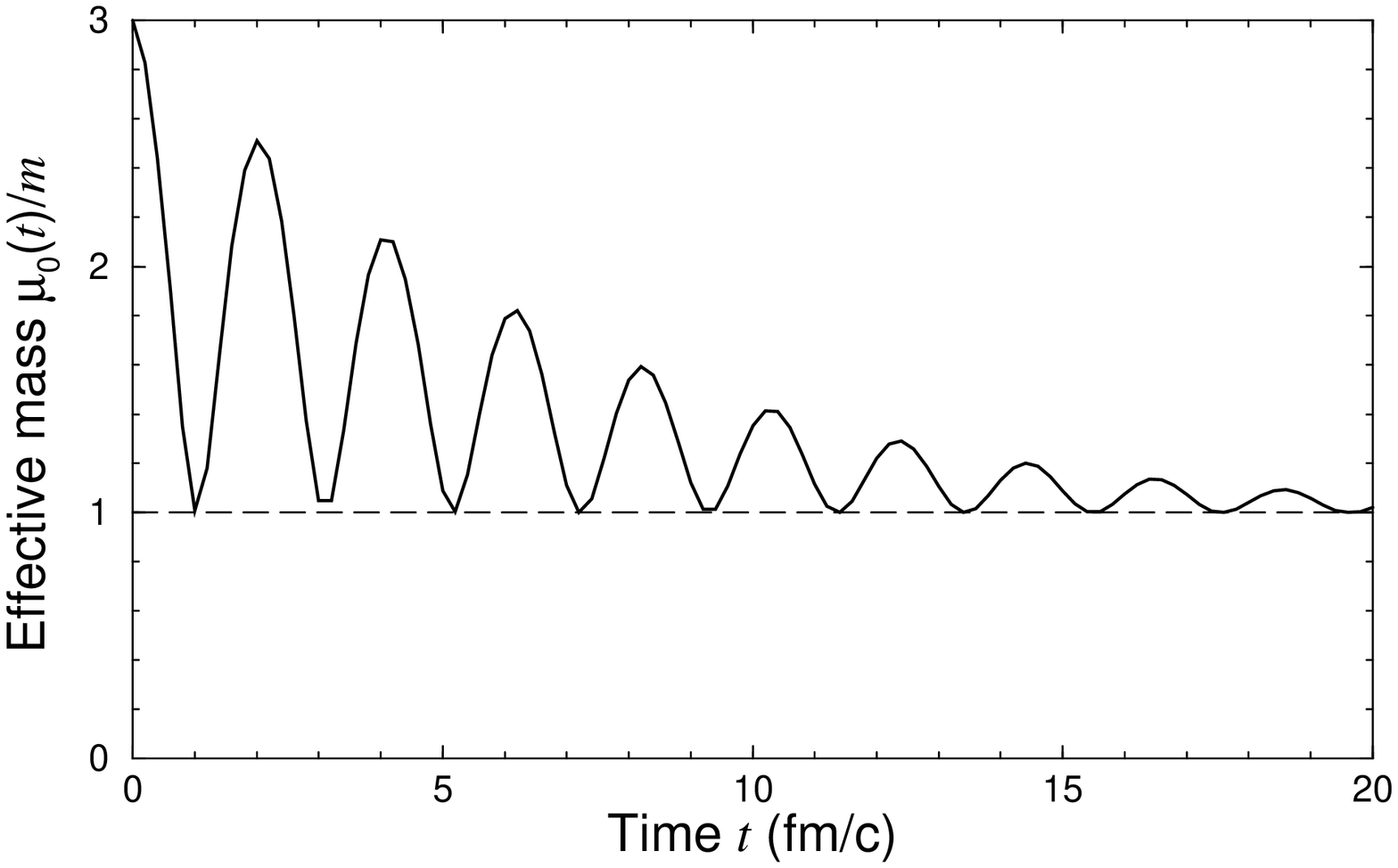,width=3.0in,angle=-90}}
\caption{
The time dependence of the bulk value of the effective mass function
used in the illustration, $\mu_0(t)$.
It has a rough correspondence with typical mass functions
resulting from dynamical DCC studies with the linear $\sigma$ model
\protect\cite{JR:PRL77,JR:PRC61}.
}\label{f:mu0}
\end{figure}

\begin{figure}[t]
\centerline{\psfig{figure=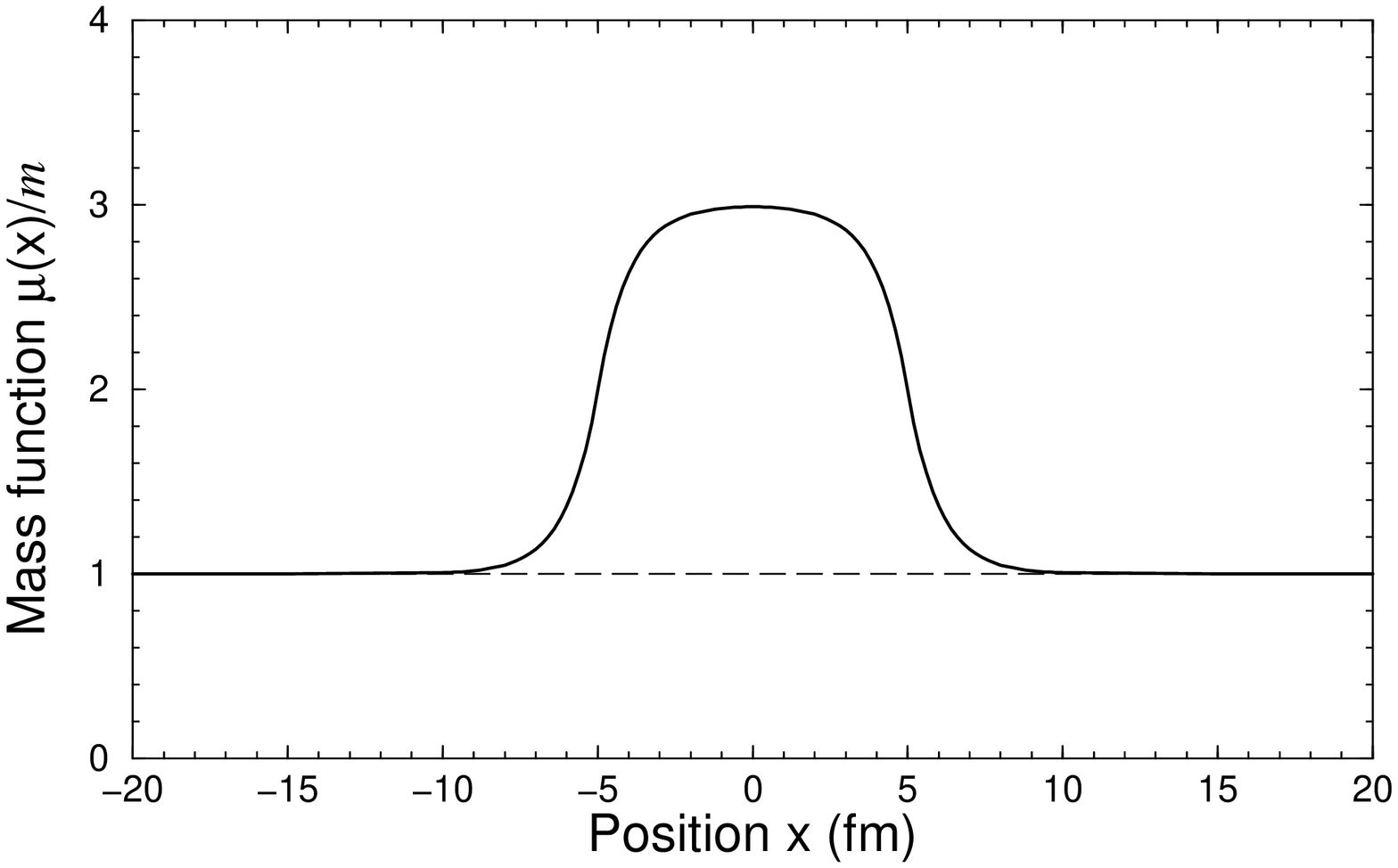,width=3.0in,angle=-90}}
\caption{
The initial mass function used in the illustration, $\mu(x,t=0)$.
It is obtained by smearing a sharp distribution on the interval
$|x|\leq5~\fm$ with an exponential having a width of one fermi.
The mass excess $\mu^2(x,t)-m^2$ is assumed to retain its spatial profile
while the overall magnitude scales as the bulk value $\mu_0(t)$
given in Fig.~\ref{f:mu0}.
}\label{f:mux}
\end{figure}

In order to demonstrate the practical applicability of the described treatment,
and at the same time illustrate the importance of the quantum fluctuations,
we consider a one-dimensional system having the most important features
expected in actual scenarios.
In order to emulate the finite transverse extension
of systems generated in high-energy heavy-ion collisions,
we assume that the initial mass function exceeds the free value $m$
within only a limited region, $|x|\leq5~\fm$.
We furthermore assume that $\mu^2(x)-m^2$ maintains this spatial profile
in the course of time,
while the overall magnitude scales in proportion to its bulk value, $\mu_0(t)$.
Fig.~\ref{f:mu0} shows the time dependence of this latter quantity,
while Fig.~\ref{f:mux} depicts the initial spatial profile of
the mass function, $\mu(x,0)$.

Given the particular mass function $\mu(x,t)$,
the mode functions $F^\lambda(x,t)$ and $G^\lambda(x,t)$
are then determined numerically by solving the equation of motion (\ref{EoM})
for each canonical mode $\lambda$.
Since the mass function remains positive throughout
(in fact it is never below the free mass $m$),
it is possible to analyze the system in terms of the adiabatic
quasiparticle modes at any time.

\begin{figure}[t]
\centerline{\psfig{figure=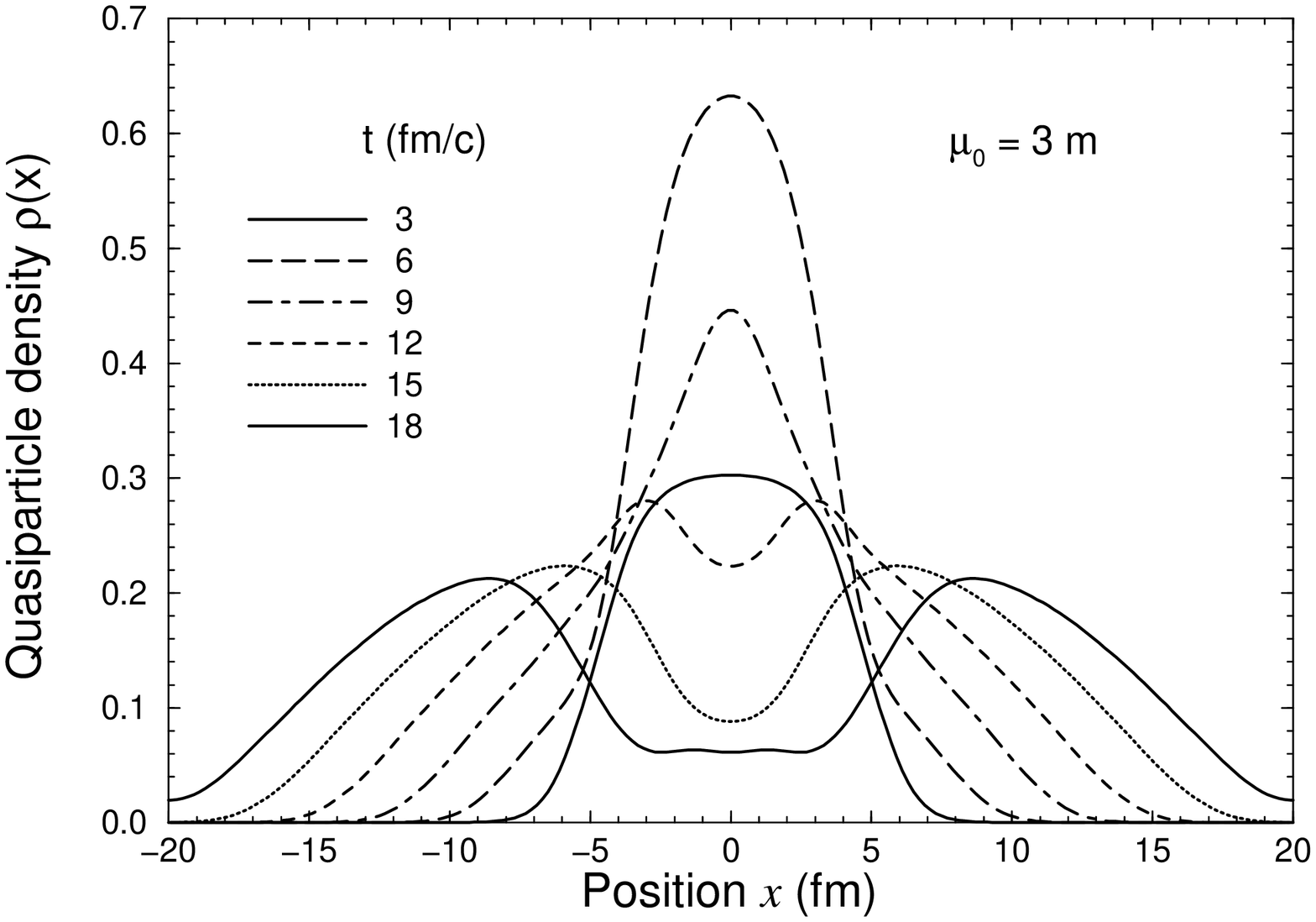,width=3.0in,angle=-90}}
\caption{
The spatial quasiparticle density at successive times $t$ (as indicated),
starting at $t=0$ from the correlated vacuum associated with
the mass function shown in Fig.~\ref{f:mux}.
}\label{f:Rvac}
\end{figure}

\begin{figure}[t]
\centerline{\psfig{figure=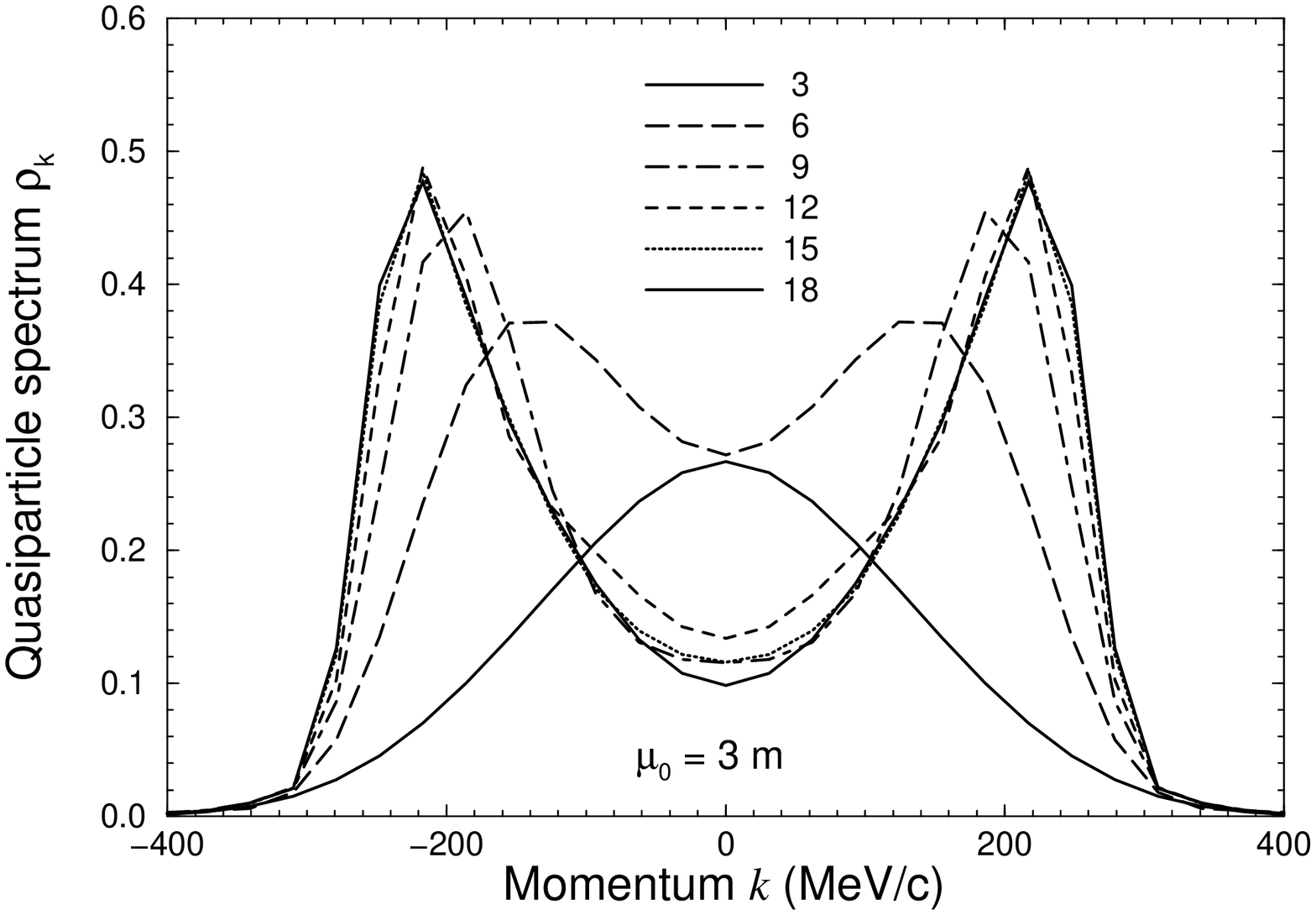,width=3.0in,angle=-90}}
\caption{
The spectral distribution of the quasiparticles at successive times
(indicated in $\fm/c$)
starting at $t=0$ from the vacuum associated with
the mass function shown in Fig.~\ref{f:mux}.
}\label{f:Qvac}
\end{figure}

The contribution from the quantum fluctuations
to the spatial quasiparticle density at various times during the evolution,
is shown in Fig.~\ref{f:Rvac}
and the associated spectral distribution is shown in Fig.~\ref{f:Qvac},
\beqar\label{obs}
\bar{\rho}^{\rm qu}(x,t)&=& \bar{\rho}^{\rm qu}(x,x;t) =
\sum_L |{\cal V}^L(x,t)|^2\ ,\\
\bar{\rho}^{\rm qu}_k(t) &=& \bar{\rho}^{\rm qu}_{kk}(t) =
\sum_L |{\cal V}^L_k(t)|^2\ .
\eeqar
These quantities show what would result
if there were no excitation at all in the initial state
(corresponding to zero temperature).
Any actual agitations in the initial state
would then lead to additional contributions.

At the initial time, $t=t_i$, the quantities $\bar{\rho}^{\rm qu}(x)$
and $\bar{\rho}^{\rm qu}_k$ vanish by design.
But soon, as the effective mass experiences its first rapid decrease,
energy stored in the high initial mass is liberated
and a rapidly rising quasiparticle density appears
with a spatial profile resembling that of the mass excess 
$\delta\mu(x)=\mu(x)-m$.
Part of these quasiparticles are reabsorbed when the mass again increases
and this cycle is repeated in a less noticable manner,
as the mass oscillations proceed with an ever decreasing amplitude.
The quasiparticles produced by this mechanism
have a rather structureless spectrum of approximately gaussian form.

Meanwhile,
the regularity of the temporal oscillations in the mass function
begins to manifest itself and, at somewhat later times,
the spectral profile exhibits a visible enhancement
of the modes near the resonance frequency,
$\omega_{\rm res}\approx\half\omega_\sigma=300~\MeV$,
corresponding to a momentum of $k_{\rm res}\approx266~\MeV/c$.
As a consequence,
the quasiparticle gas begins to resemble two countermoving beams
with the corresponding flow velocities centered around
$v_{\rm flow}\approx\pm k_{\rm res}/m$.
The two corresponding outwards bursts are clearly visible in Fig.~\ref{f:Rvac}
at late times when the mass function has largely relaxed to its free form
and the production processes have subsided.

Traces of these characteristic effects
might be observable in actual collision experiments,
such as those underway at RHIC.
More extensive calculations would clearly be needed
for a quantitative assessment of this prospect.

\section{Discussion}
\label{conclude}

In the present paper we have sketched how
the non-uniform and rapidly evolving scenarios
of interest  in connection with disoriented chiral condensates
may be addressed with real-time non-equilibrium quantum-field theory
within the mean-field approximation,
in which the interactions are encoded into an effective mass function.
For an arbitrary form of the mass function,
the strategy is to first obtain the time evolution operator,
which depends only on $\mu^2(\x,t)$,
and then evaluate the observables
resulting from any particular initial quantum state $|\ti\rangle$.
This makes it economical to perform averages over ensembles of initial states
having similar mass functions.

When the eigenvalues of the mass matrix are positive,
$\Omega_\K^2>0$,
it is particularly instructive to consider quasiparticle excitations
defined relative to the adiabatic eigenbasis.
But it should be noted that even if the system enters
the classically forbidden region where the quasiparticles cannot be defined,
the dynamical propagation of the system works without modification.

A first application of the developed treatment has been made
to a system with a non-trivial mass function
depending on both time and space.
In addition to demonstrating the practicality of the approach,
the example serves to illustrate the quantitative importance
of including the quantum fluctuations in the dynamical treatment.

The approach assumes that the effective mass function is given.
For the time being,
an approximate form of the mass function for a given scenario
can be obtained on the basis of semiclassical simulations \cite{JR:PRC61}.
Though not fully satisfactory,
this approach may actually be reasonably accurate,
since the quantum treatment,
while having a large effect on specific signals,
is expected to have a relatively small overall effect on the effective mass.
Nevertheless,
an interesting and more formal task, beyond scope of present study,
is posed by the question of how to best determine the effective mass function
selfconsistently with the evolution of the quantum many-body state.

As noted above,
the computational challenge posed by the treatment
is not prohibitively larger than that met at the classical level,
since it merely requires (repeated) solution of the same field equation
(\ref{EoM}).
The treatment may thus be applied to scenarios similar to those
addressed at the classical level.
This presents an obvious practical task
which should yield more quantitative information
on the observable signals of the expected chiral dynamics.

\section*{Acknowledgements}
The author is pleased to acknowledge helpful discussions with
D.\ Boyanovsky, S.J.\ Jeon, and E.\ Mottola.
This work was supported by the Director, Office of Energy Research,
Office of High Energy and Nuclear Physics,
Nuclear Physics Division of the U.S. Department of Energy,
under Contract DE-AC03-76SF00098.




\vfill{\small \noindent{\em LBNL-46265 \hfill Physical Review C62}}

                        \end{document}